# Initialization of a nuclear spin system over the quantum Hall regime for quantum information processing

**R G Mani,[*] W B Johnson,[‡] and V. Narayanamurti[*]**

[*]Harvard University, Gordon McKay Laboratory of Applied Science, 9 Oxford Street, Cambridge, MA 02138, U.S.A.

[‡]Laboratory for Physical Science, University of Maryland, College Park, MD 20740, U.S.A.

**Abstract.** The application of the quantum mechanical properties of physical systems to realize novel computational schemes and innovative device functions have been topics of recent interest. Proposals for associated devices are to be found in diverse branches of physics. Here, we are concerned with the experimental realization of some elements needed for quantum information processing using nuclear spin immersed in a confined electronic system in the quantum Hall regime. Thus, we follow a spin-handling approach that (a) uses the Overhauser effect in the quantum Hall regime to realize a large nuclear polarization at relatively high temperatures, (b) detects the nuclear spin state by measuring the influence of the associated magnetic field on Electron Spin Resonance, and (c) seeks to apply the electronic spin exciton as the spin transfer mechanism. Some measurements examining the viability of this approach are shown, and the utility of the approach for initializing a nuclear spin system at a relatively high temperature is pointed out.

## 1. Introduction

Quantum computers are machines that exploit the quantum mechanical properties of physical systems to realize an exponential speed-up in problem solving capability compared to existing computers.[1] According to DiVincenzo,[2] realization of such a machine requires five essential components: (i) the machine should have a collection of bits, (ii) it should be possible to set all the memory bits to 0 before the start of each computation, (iii) the error rate should be sufficiently low, (iv) it must be possible to perform elementary logic operations between pairs of bits, and (v) reliable output of the final result should be possible.[2] Existing approaches that include these essential features are to be found in diverse fields such as atomic physics, quantum optics, superconductivity, nuclear magnetic resonance, and solid state physics.[2] We are concerned here with the experimental realization of some elements needed for quantum information processing using nuclear spin immersed in a confined electronic system in the quantum Hall regime.[3] Some characteristics of this approach include the application of the Overhauser effect for dynamic nuclear polarization,[4] spin detection using relatively simple electrical measurements, and the utilization of a semiconductor host.

## 2. Background and Approach

Spin based quantum computers in low dimensional semiconductors may be broadly grouped into nuclear spin quantum computers and electron spin quantum computers.[2] Among these are: (a) An early theoretical nuclear spin quantum computing proposal by Privman, Vagner, and Kventsal which identified nuclear spins in quantum Hall systems as qubits,[5] and exploited an impurity mediated two qubit interaction as the medium for a universal quantum logic gate.[5,6] In this scheme,[5] measurement relied upon detecting the spin polarization in currents directed across qubit replicas, while logic control relied upon precise, local modification of the impurity state. Notably, this theoretical proposal raised a number of spin measurement and control issues that needed to be addressed by experimenters. (b) A scheme by Kane for nuclear spin quantum computing in the canonical semiconductor Si,[7] which identified the possibility of (i) gate controlled single nuclear spin rotation through the regulation of the overlap between the donor electron and the nucleus in P-doped Silicon, (ii) gate controlled two qubit interaction by electronic spin exchange, and (iii) state readout by capacitive detection of the current associated with charge transfer between a pair of donors depending upon the state of the nuclear spin qubits.[7] Here, device implementation requires precise atomic control in the placement of donor atoms within the Si host, and the subsequent registry of narrow gates for single donor control. (c) A plan by Loss and DiVincenzo for electron spin quantum computing using single electron quantum dots, which invokes ferromagnetic quantum dots for state preparation, gate controlled dot overlap for transient two qubit interactions, and qubit state measurement by gate controlled tunneling into a supercooled paramagnetic quantum dot or spin valve controlled tunneling into a single electron quantum dot electrometer.[8] (d) A method for donor atom electron spin quantum computing in Si/Ge through g-factor engineering.[9] This approach applies gate control to tune the donor electron resonance frequency, and realizes two qubit interaction through the gate controlled overlap of the electronic wavefunction on neighboring sites. State readout requires that the charge state of the donor ($D^0$, $D^+$, $D^-$) modulate a 'channel' current in a conducting plane.

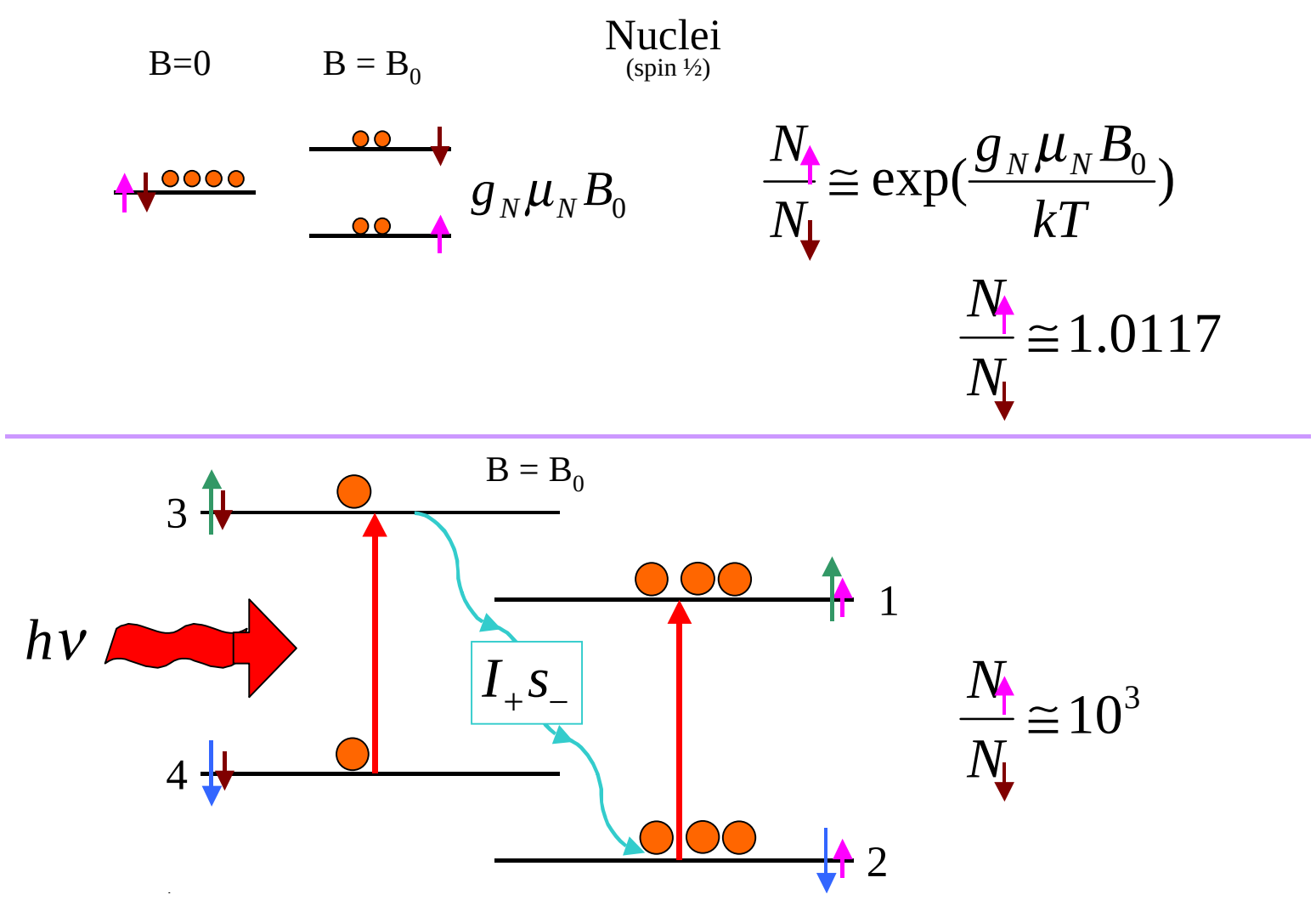

Fig. 1) (Top) A nuclear spin system exhibiting a Nuclear Magnetic Resonance frequency of 7.3 MHz/T shows nearly equal occupancy of the up and down spin levels at 0.3 K and 10 Tesla. (Bottom) The Overhauser effect can be used to massively boost the nuclear spin polarization under the same conditions.

In evaluating the relative merits of electron spin quantum computing *vs.* those of nuclear spin quantum computing, the phase coherence time, $t_\phi$, and the switching time, $t_{switch}$, for carrying a single qubit operation need to be examined and, in particular, the ratio $t_\phi/t_{switch}$ which determines the number of operations that can be carried out in a time period $t_\phi$, needs to be evaluated.[1,2] Optimistic estimates indicate that as many as $10^7$ operations can be carried out in nuclear spin QC's versus only $10^4$ operations in electron spin quantum computers. This point, and the possible long nuclear spin coherence time (up to $10^4$ seconds),[1] identify a special simple advantage for nuclear spin quantum computing: nuclear spins allow more time for the experimenter to carry out qubit manipulations. However, this advantage for nuclear spin quantum computing is countered by the shortcoming that comes with the small energy scales of nuclear spins, namely, the need to work at extremely low temperatures, T, and high magnetic fields, B.

The hybrid scheme examined here aims to overcome this shortcoming of nuclear spin based information systems by applying the Overhauser effect in the quantum Hall regime to realize large nuclear polarization at relatively high temperatures. Thus, this approach combines the favorable time scales of nuclear spins with the large energy scales of the electronic spin system. The nuclear spin state is also to be detected by examining the influence of the associated magnetic field on Electron Spin Resonance (ESR).[10] Finally, an electronic spin exciton is to be tested as a possible mobile spin transfer mechanism that could be useful for the realization of logic.[11]

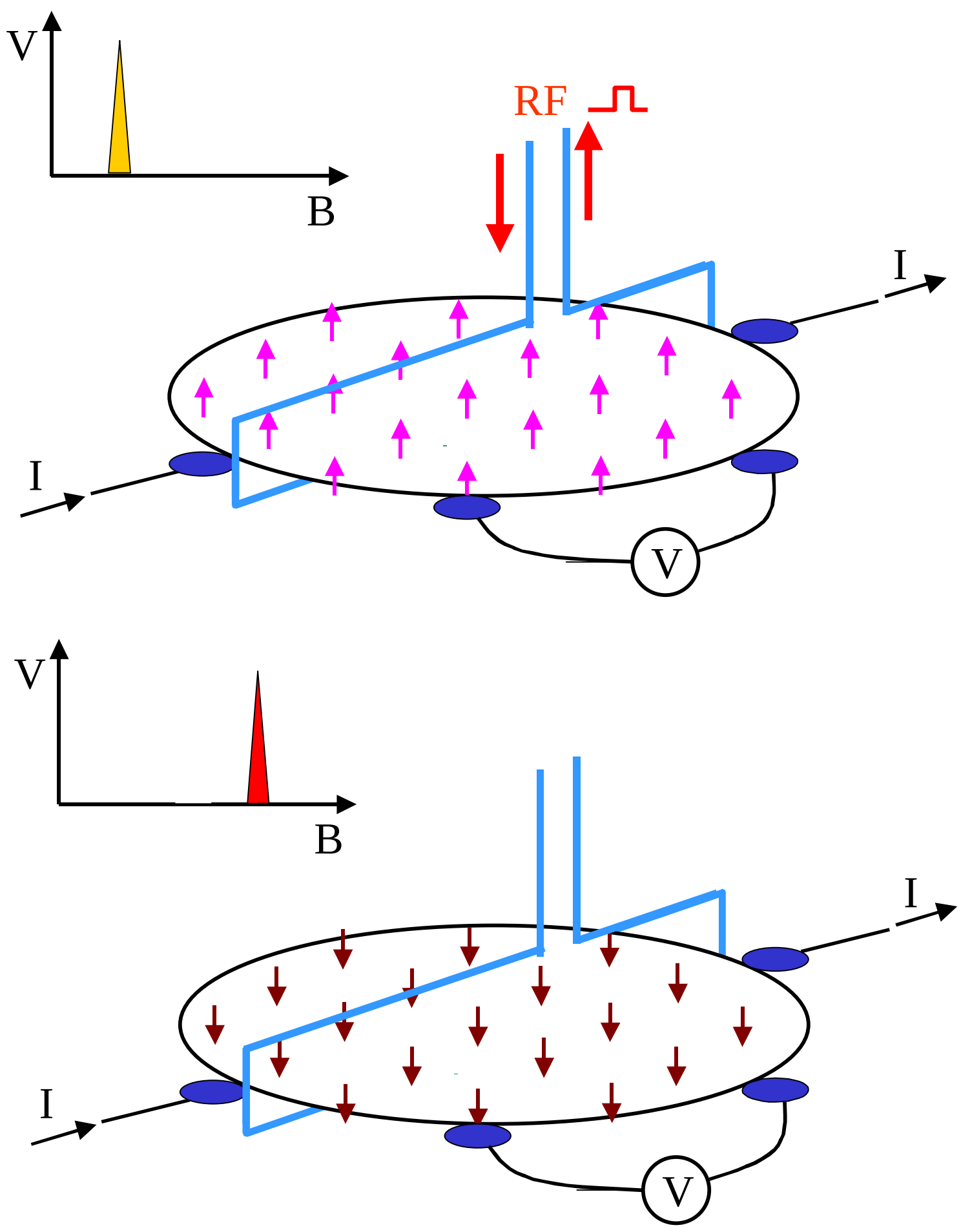

Fig. 2) A nuclear spin system that has been polarized using the Overhauser effect (Fig. 1 (bottom)) can be subjected to spin rotation by applying rf-pulses. The insets at the top and bottom illustrate the shift in the ESR line as a function of nuclear spin orientation

The effect of a small nuclear spin energy is illustrated in Fig. 1(top), which shows that a nuclear spin system with a spin flip NMR frequency of, say, 7.3 MHz/Tesla, at easily attained experimental parameters such as T = 0.3 K and B = 10 Tesla, shows nearly equal population of the up and the down spin levels in thermal equilibrium. Thus, an attempt to write information onto such a nuclear spin system, would yield unreliable results because the initial state of nuclei will not be known, a priori, with certainty. If, however, one uses the Overhauser effect to polarize the nuclei via the hyperfine flip-flop interaction as shown in Fig. 1 (bottom), then the nuclear polarization ratio can be increased by a factor of nearly a thousand for the same experimental parameters, and this helps to realize a reliable initial state, one where different spins evolve to the same state with the same operation.

The Overhauser effect involves the hyperfine interaction which couples the electronic and the nuclear spins.[4] Here, the spin coupling term (I.S) contributes the diagonal term towards the energy, which shows a four level structure (see Fig. 1(bottom)) depending on the z-component of the electronic and nuclear spins. A saturation of the electron spin resonance by the application of intense microwave radiation equalizes the population of the levels. Then, spin relaxation via the off-diagonal flip-flop term $I_+S_-$ serves as a channel for establishing a steady state, see Fig. 1 (bottom), leading to a large population difference between the nuclear spin states. Here, the Boltzmann factor describing steady state is determined mostly by the electronic spin-flip energy, which is much larger than the nuclear spin flip energy. Thus, in this scheme, the nuclei are more likely to be found in a known energy state, at a high temperature.[4]

After initialization, the nuclei can be rotated using rf- pulses as illustrated in Fig. 2. And, the result can be tracked by measuring the ESR line field shift using the four terminal resistance. Here, the electron spin resonance line shifts in proportion to the nuclear magnetic field because the spin polarized nuclei produce an effective magnetic field $B_N$ that modifies the electronic spin splitting according to the relation $E_S = g\mu_B(B + B_N)$, where $g\mu_B B$ is the bare electronic spin splitting.[10] Thus, in addition to a method for state initialization, this approach also includes a method for the readout of the nuclear spin state, namely, through a measurement of the shift of the ESR line.

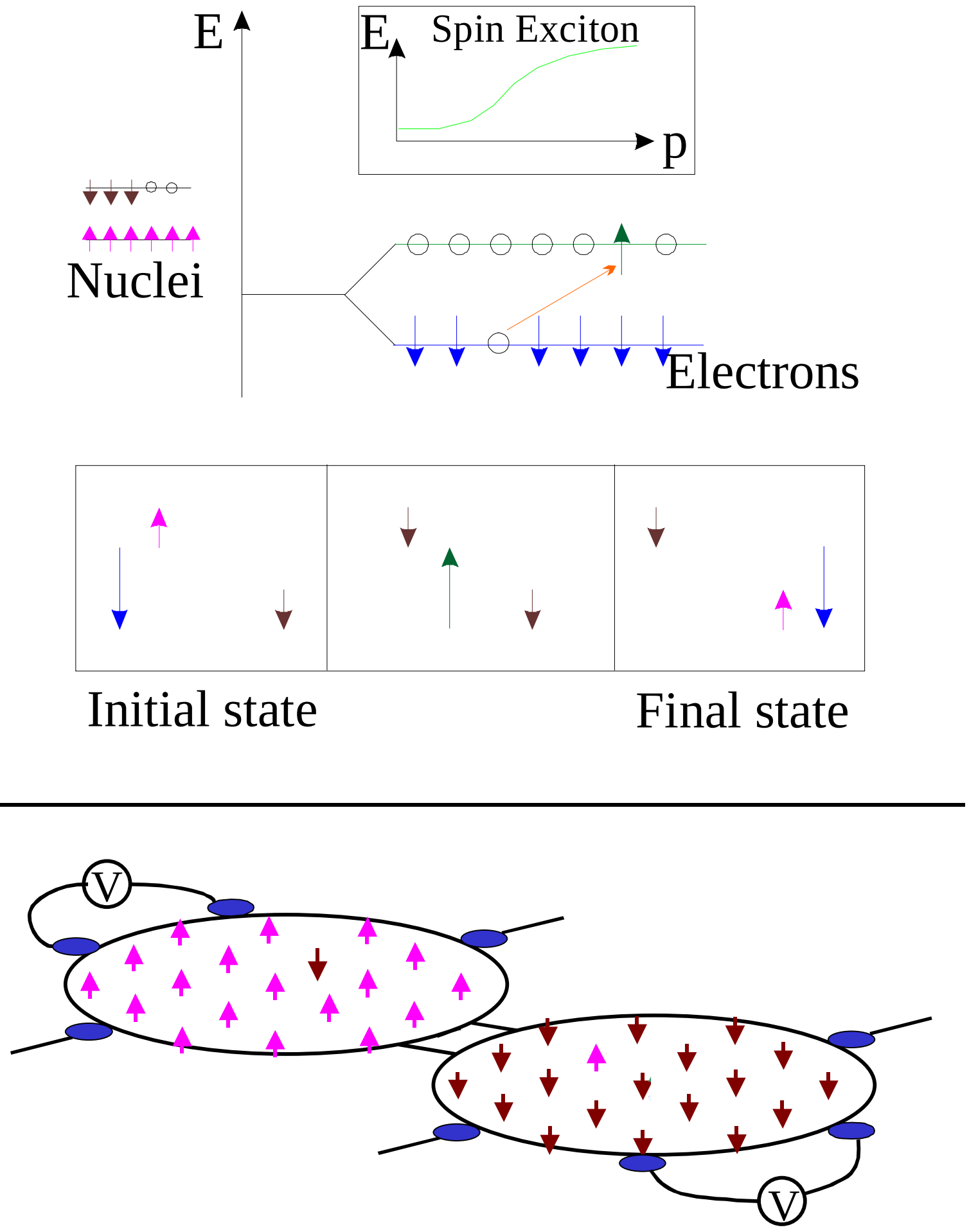

Fig. 3) (Top) A quantum Hall system at odd filling factors can spontaneously form electronic spin excitons (ref. 11), which can serve as a medium (see Center) for transporting nuclear spin polarization. (Bottom) A pair of nuclear spin polarized quantum dots viewed after a spin exciton has transported spin polarization from one dot to another.

A third necessary component for quantum information processing is an interaction mechanism for coupling a pair of qubits. It has been suggested that nearly any pair interaction can serve to realize this function.[1,2] We will attempt to apply the spin exciton mechanism suggested for quantum Hall systems (See Fig. 3).[11] A spin exciton is produced by flipping an electronic spin from the lower energy 'down' state to the higher energy 'up' state. Then, the hole thus created in the 'down' state binds via the Coulomb interaction with the (promoted) electron in the high energy ‘up’ state to form a ‘spin exciton,’ and this exciton propagates within the specimen.[11] For example, a spin up nucleus immersed in a sea of spin down electrons, can trade its spin with a nearby spin down electron. The flipped electron or spin exciton can now propagate to a distant nucleus and exchange spin with it, as illustrated in Fig. 3(center).[11] The main features in this scheme are therefore (i) reading out the nuclear spin state by detecting the influence of the nuclear spin magnetic field on the electron spin resonance, (ii) using electrical techniques for electron- and nuclear- spin resonance detection, (iii) endowing nuclear spins with the relatively large energy scales of electronic spins using the hyperfine interaction and the Overhauser effect (Fig. 1), (iv) applying radio frequency pulses and electric control to realize selective nuclear spin rotation (Fig. 2), and (v) utilizing semiconductor technology to realize a plurality of devices on a single chip, which can be independently controlled and separately measured by electrical means.

## 3. EXPERIMENT

Thus far, the experiments have examined the electrical response of GaAs/AlGaAs heterostructure devices under simultaneous microwave ($27 \leq f \leq 60$ GHz) and radio-frequency ($30 \leq f \leq 100$ MHz) excitation, over the range $1.2 \leq T \leq 4.2$ K and $B \leq 12$ Tesla, in order to confirm the mechanisms of interest.

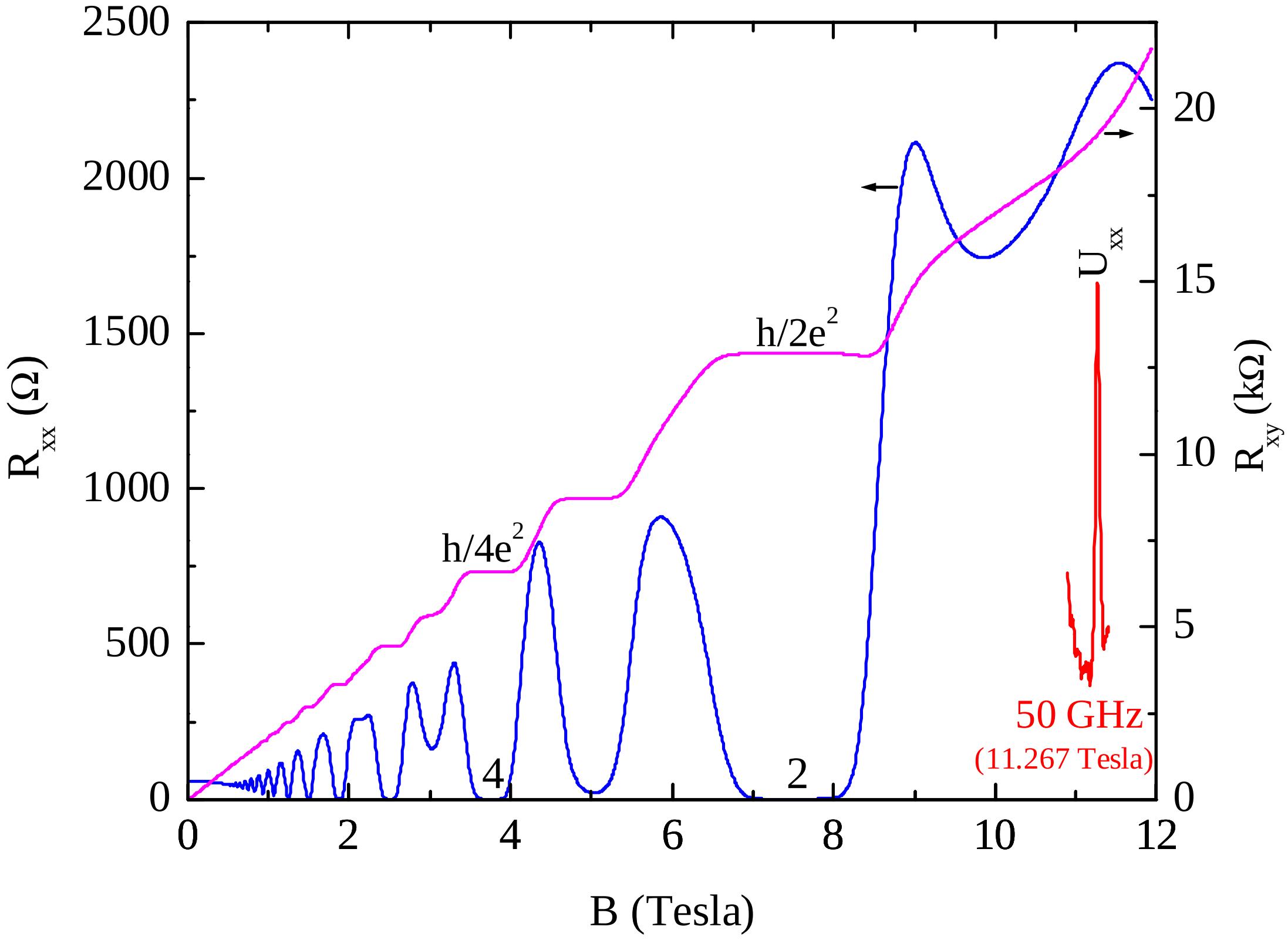

Fig. 4) Transport characteristics of a GaAs/AlGaAs device under microwave excitation over the quantum Hall regime. The microwave induced voltage $U_{xx}$ shows Electron Spin Resonance.

Figure 4 shows the transport response of a device under microwave excitation at 50 GHz. Here, the diagonal resistance $R_{xx}$ exhibits Shubnikov-deHaas oscillations with increasing B and vanishing $R_{xx}$ over wide B intervals, while the Hall resistance $R_{xy}$ exhibits a linear-in-B increase and quantization (in units of $h/e^2$), as expected for quantum Hall effect.[3] Also shown in the figure is the microwave induced ESR in the vicinity of filling factor $\nu = 1$ that is detected using a dual lock-in technique. Here, one observes a large sharp resonant response in the data, which demonstrates a good signal to noise ratio. Figure 5 summarizes ESR measurements at a number of microwave frequencies both in the vicinity of filling factor $\nu = 1$, with the sample perpendicular to the magnetic field, i.e., $\theta = 0^0$, and in the vicinity of $\nu = 3$ with $\theta = 60^0$.

The experimental signature of dynamic nuclear polarization due to the Overhauser effect is shown in Fig. 6 for a quantum Hall system in the vicinity of $\nu = 3$ at $\theta = 60^0$. Here, the microwave induced voltage shows a narrow resonance on the up-sweep of B, signifying ESR. However, on the B down-sweep, one observes large hysteresis because the ESR condition is maintained over a wider range of magnetic fields as polarized nuclear spins contribute a magnetic field $B_N$ that helps maintain the ESR condition $E_S = g\mu_B(B + B_N)$ at fields below where one normally expects ESR. These data demonstrate the good signal-to-noise ratio and the good sensitivity to nuclear spin polarization which can be obtained in such experiments.

The suggested role for nuclei in such hysteretic transport (Fig. 6) can be confirmed by attempting to change the nuclear polarization by NMR over the hysteretic region.[10] For this purpose, measurements were carried out with simultaneous microwave and radio-frequency (rf) irradiation of the specimen, with the rf chosen to coincide with the nuclear magnetic resonance frequency of the host nuclei, i.e., Ga or As. The results of such experiments, which sought to detect NMR of the $Ga^{69}$ nuclei are illustrated in Fig. 7. The inset of this figure shows the downsweep electrical response of the specimen with rf radiation at 85, 85.5, and 86 MHz, respectively. The observation of resonance in Fig. 7 appears to confirm that dynamic nuclear polarization is responsible for the observed hysteretic transport in Fig. 6.

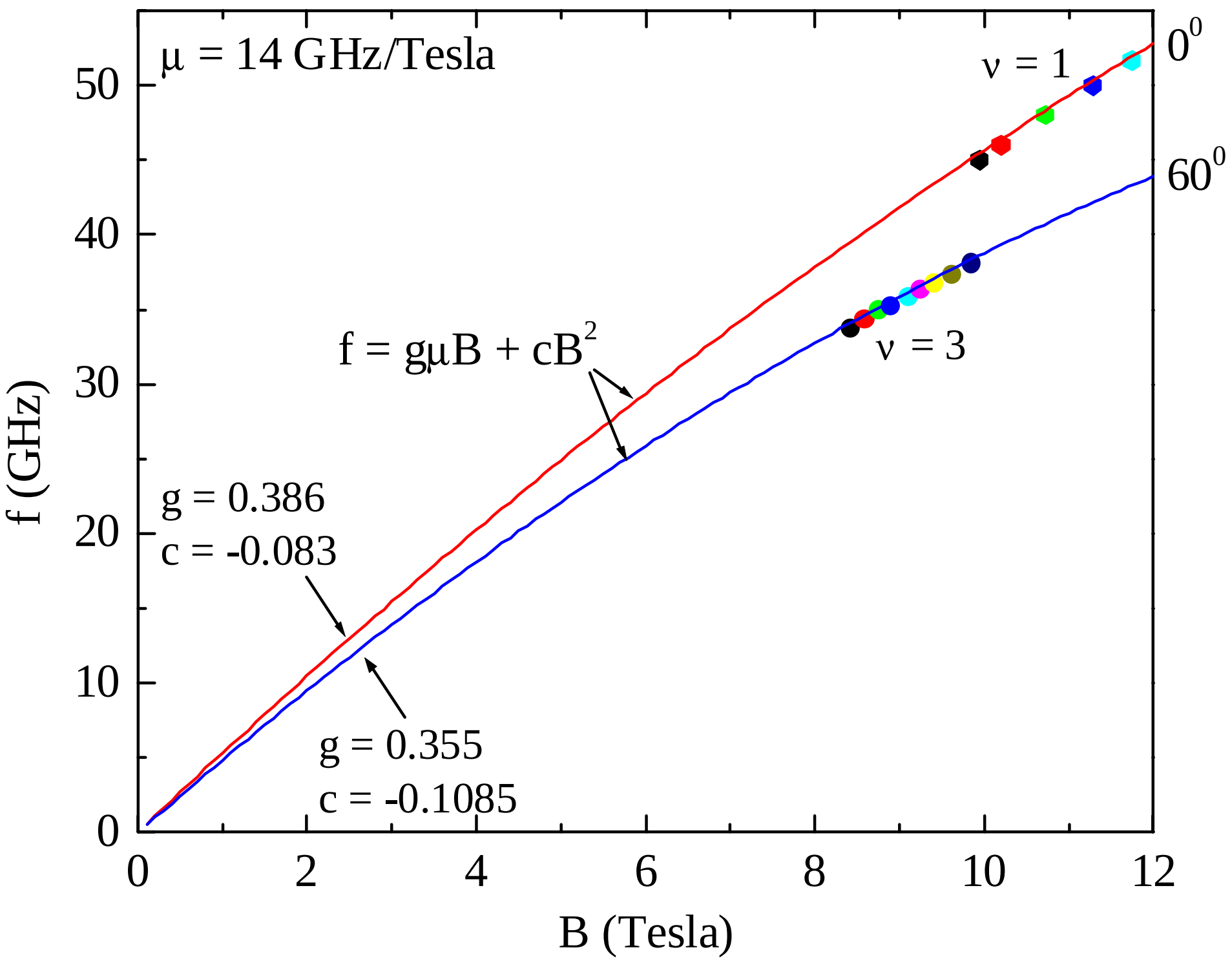

Fig. 5) Electron Spin Resonance frequencies as a function of the magnetic field in a GaAs/AlGaAs device at 1.3 K in the vicinity of filling factor $\nu = 1$ and $\nu = 3$.

The magnitude of the nuclear spin polarization in the experiment of Fig. 6 can be estimated by taking into account the contributions of the elemental components $^{69}$Ga (I = 3/2) , $^{71}$Ga (I = 3/2), and $^{75}$As(I = 3/2), which provide field shifts $B_N^{69\ Ga}$ = -0.91 T $<I^{69\ Ga}>$, $B_N^{71\ Ga}$ = -0.78 T$<I^{71\ Ga}>$, and $B_N^{75\ As}$ = -1.84 T$<I^{75\ As}>$, respectively. Thus, in this instance (Fig. 6), the data suggest that a little less than 10% of the nuclei have been polarized at 1.3 K. Clearly, this is a substantial increase over the expected nuclear polarization at the same temperature based on thermal equilibrium estimates (cf. Fig. 1) and this feature confirms the operation of the Overhauser mechanism in these experiments.

Such studies also suggest that the nuclear spin polarization depends sensitively upon the temperature, the sweep rate, and the microwave intensity. Although the detailed dependence of the field shift upon these parameters continues to be under study, the measurements indicate that this is a promising approach for initializing the nuclear spin system at a relatively high temperature, compared to nuclear spin flip energy scales. Once the nuclei have been initialized, then the long lifetimes of nuclei might be exploited

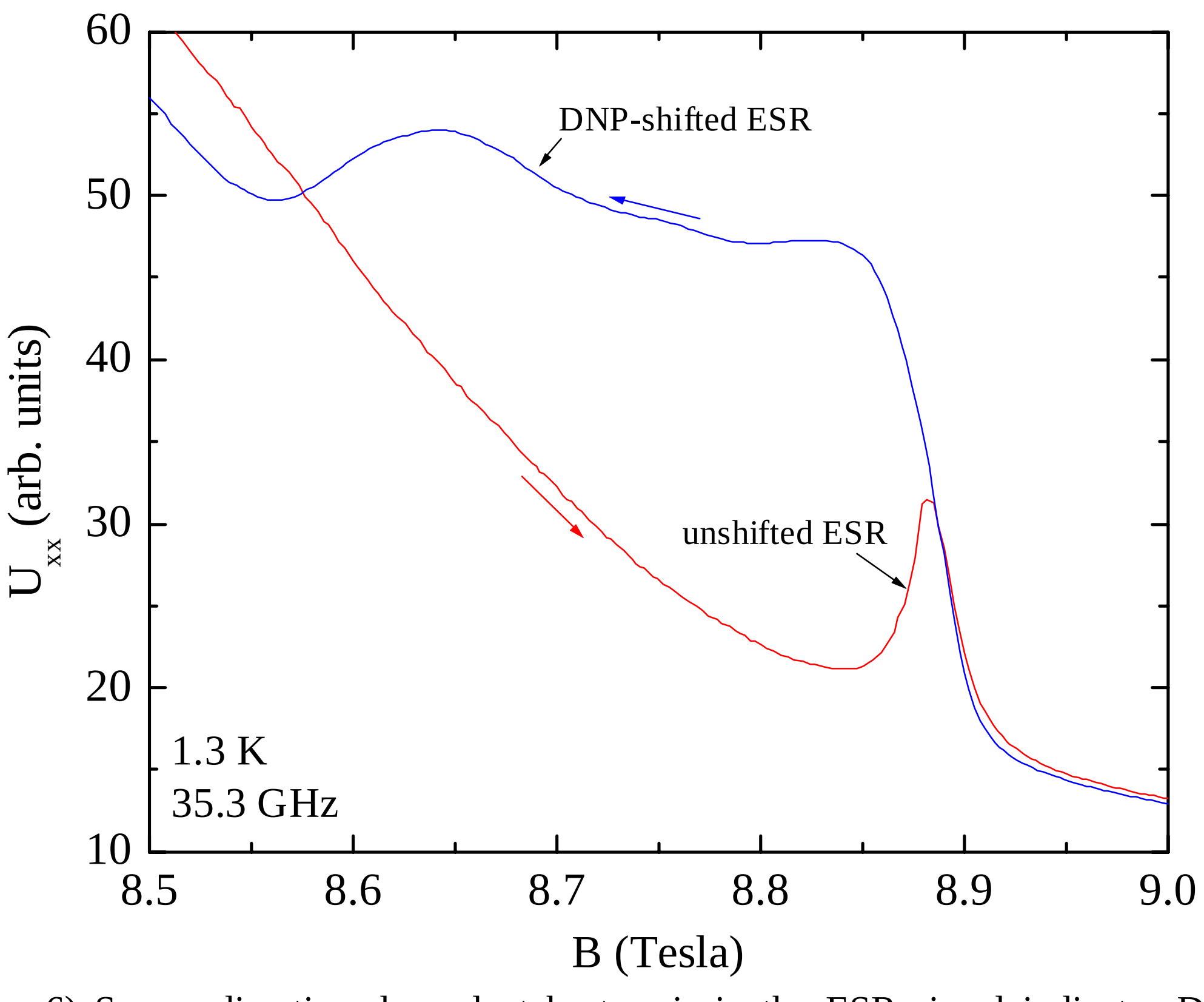

Fig. 6) Sweep direction dependent hysteresis in the ESR signal indicates Dynamic Nuclear Polarization (DNP) via the Overhauser effect. The tilt angle is $60^0$.

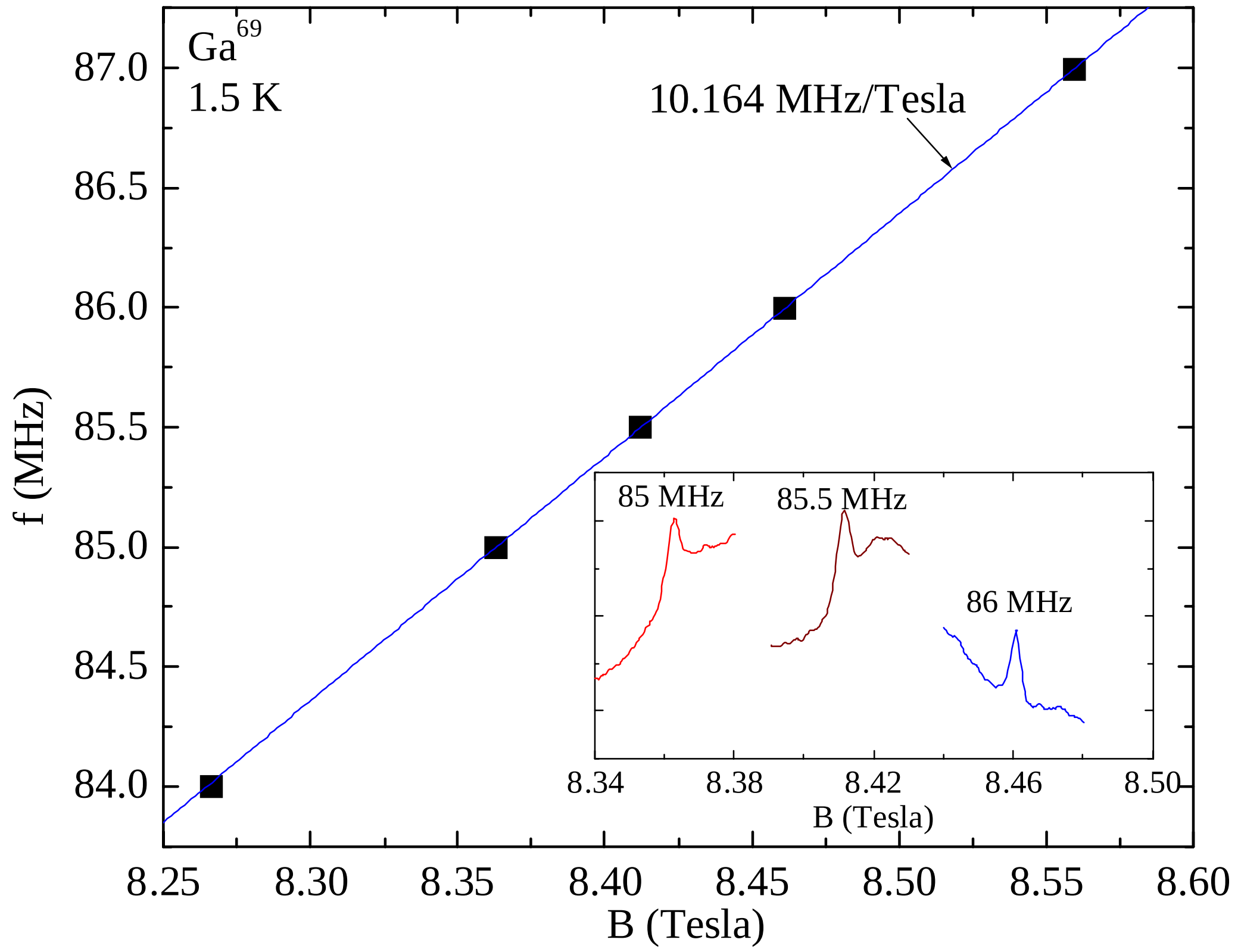

Figure 7) (inset) The electrically detected Nuclear Magnetic Resonance of $Ga^{69}$ nuclei in a GaAs/AlGaAs device illuminated with 34.5 GHz microwave radiation at a tilt angle of $60^0$ and radio frequency radiation as indicated. The main panel shows the linear shift of the NMR frequency with increasing magnetic field.

to carry out, at a deliberate pace, pulse rf based nuclear spin manipulations, before state readout.

## 5. Acknowledgements

We acknowledge stimulating discussions with J. H. Smet and K. von Klitzing, and measurement visits to the Max-Planck-Institut FkF (Stuttgart). Thanks to V. Privman, D. Mozyrsky, and I. Vagner for theoretical support. Finally, we would like to acknowledge M. Dobers for sharing his expertise relating to ESR and NMR in quantum Hall systems. This work has been supported by the ARO, SRC, DFG, and BMBF.